\documentclass[11pt,letterpaper]{article}%
\usepackage{amsmath}
\usepackage{epsfig}
\usepackage{graphicx}%
\usepackage{amsfonts}%
\usepackage{amssymb}
\begin{document}
%
%TCIMACRO{\TeXButton{titlepage}{\begin{titlepage}
%\begin{flushright}
%UCL-IPT-04-02
%\end{flushright}
%\vspace*{30mm}
%\begin{center}
%\huge{QCD anomalies in hadronic weak decays}
%\end{center}
%\vspace*{10mm}
%\begin{center}
%\Large{Jean-Marc G\'{e}rard\footnote{gerard@fyma.ucl.ac.be}
%and St\'{e}phanie Trine\footnote{trine@fyma.ucl.ac.be}}
%\end{center}
%\vspace*{5mm}
%\begin{center}
%Institut de Physique Th\'{e}orique, Universit\'{e} catholique de Louvain,
%\\Chemin du Cyclotron, 2, B-1348, Louvain-la-Neuve, Belgium
%\end{center}
%\vspace*{5mm}
%\begin{center}
%February 16, 2004
%\end{center}
%\vspace*{10mm}
%\begin{abstract}
%We consider the flavour-changing operators associated with the strong axial and trace
%anomalies. Their short-distance generation through penguin-like diagrams is obtained
%within the QCD external field formalism. Standard-model operator evolution exhibits a
%suppression of anomalous effects in K and B hadronic weak decays. A genuine set of
%dimension-eight $\Delta S=1$ operators is also displayed.
%\end{abstract}
%\end{titlepage}}}%
%BeginExpansion
\begin{titlepage}
\begin{flushright}
UCL-IPT-04-02
\end{flushright}
\vspace*{30mm}
\begin{center}
\huge{QCD anomalies in hadronic weak decays}
\end{center}
\vspace*{10mm}
\begin{center}
\Large{J.-M. G\'{e}rard\footnote{gerard@fyma.ucl.ac.be}
and S. Trine\footnote{trine@fyma.ucl.ac.be}}
\end{center}
\vspace*{5mm}
\begin{center}
Institut de Physique Th\'{e}orique, Universit\'{e} catholique de Louvain,
\\Chemin du Cyclotron, 2, B-1348, Louvain-la-Neuve, Belgium
\end{center}
\vspace*{5mm}
\begin{center}
February 16, 2004
\end{center}
\vspace*{10mm}
\begin{abstract}
We consider the flavour-changing operators associated with the strong axial and trace
anomalies. Their short-distance generation through penguin-like diagrams is obtained
within the QCD external field formalism. Standard-model operator evolution exhibits a
suppression of anomalous effects in K and B hadronic weak decays. A genuine set of
dimension-eight $\Delta S=1$ operators is also displayed.
\end{abstract}
\end{titlepage}%
%EndExpansion

\newpage

\section{Introduction}

Broken symmetries have often proved as useful as exact symmetries to describe
the physical world. In this respect, particular attention has been paid to the
anomalies of Quantum Field Theory \cite{Bertlmann96}. Their presence in global
axial \cite{AxialAno} and scale \cite{TraceAno} symmetries of massless $QED$
and $QCD$ is notably required by the physics of light (pseudo-)scalar particle
decays into two photons \cite{Steinberger49} and hadronic transitions between
quarkonium levels \cite{NSVZ80}.

In hadronic weak decays, the $QCD$ axial and trace anomalies have already been
advocated to understand the unexpectedly large branching fractions for
$B\rightarrow\eta^{\prime}X_{s}$ \cite{AnoBtoEtap} and the $\Delta I=1/2$
selection rule in $K\rightarrow\pi\pi$ \cite{PenPiv2}\cite{Gerard01},
respectively. However, to our knowledge, no real attempt has been made so far
to generate \emph{both} pseudoscalar and scalar anomalous gluonic
configurations from short-distance $QCD$ corrections to a four-quark
flavour-changing effective interaction.

At lowest order in the momentum expansion, the local anomalous operators in
request consist of a flavour-changing quark bilinear coupled to the
colour-singlet gluon density $G^{\mu\nu}\widetilde{G}_{\mu\nu}$ in the axial
anomaly case and $G^{\mu\nu}G_{\mu\nu}$ in the trace anomaly one. At the
one-loop level, they are thus generated by gluonic penguin-like diagrams with
a heavy virtual flavour. In this paper, we infer the corresponding $QCD$
evolution of an arbitrary four-quark operator from the propagator of the heavy
intermediate flavour plunged in external gluon fields and expanded in inverse
powers of the mass. The contributions of the leading anomalous operators are
then displayed.

Due to our assumptions of a four-quark initial operator and a heavy
intermediate flavour, the method turns out to be particularly well-suited to
the estimation of short-distance charm effects in Kaon decays within the
Standard Model. Yet, the insight gained allows us to clarify the role of heavy
quark generated anomalies in $B$ decays too. Besides, the formalism developed
here is applicable to any four-quark effective interaction induced by physics
beyond the Standard Model.

A by-product of our method is the consistent derivation of a complete set of
dimension-eight operators in standard-model hadronic $K$ decays. Contrary to
ref.\cite{PenPiv1}, the central role granted here to the heavy flavour
propagator immersed in $QCD$ external fields allows us indeed to avoid a
misuse of the quark classical equations of motion.

\section{Heavy quark induced penguin-like operators}

Let us start with the computation of the one-loop penguin-like short-distance
$QCD$ corrections to the generic four-quark operator
\begin{equation}
Q=(\overline{q}_{1}\Gamma^{A}q_{h})(\overline{q}_{h}\Gamma^{B}q_{2}%
),\label{4QuarkOp}%
\end{equation}
under the assumption of a heavy intermediate flavour $q_{h}$. The symbols
$\Gamma^{A,B}$ stand for products of Dirac matrices. The quark bilinears are colour-singlets.

In the path integral formalism, these corrections may be obtained by
functional integration of the heavy mode $q_{h}$ in the presence of a
classical gluonic background \cite{Dobado}:
\begin{equation}
\exp i\int d^{4}x\mathcal{L}_{eff}=\int\mathcal{D}q_{h}\mathcal{D}\overline
{q}_{h}\exp i\int d^{4}x\left(  \overline{q}_{h}%
%TCIMACRO{\TeXButton{slash}{\! \not\!} }%
%BeginExpansion
\! \not\!
%EndExpansion
Pq_{h}-M\,\overline{q}_{h}q_{h}+g_{w}Q\right) \label{GaussianInt}%
\end{equation}
with $g_{w}$, the effective coupling associated with the weak operator $Q$,
and $M$, the heavy flavour mass. The covariant derivative acting on spinor
fields is defined by
\begin{equation}
P_{\mu}=i\partial_{\mu}+g_{s}A_{\mu}\label{CovDer}%
\end{equation}
with $A_{\mu}=A_{\mu}^{a}\lambda^{a}/2$, the external gluon field and $g_{s}$,
the $QCD$ coupling constant. The Gell-Mann matrices $\lambda^{a}$ are
normalized such that $tr\lambda^{a}\lambda^{b}=2\delta^{ab}$ $(a,b=1,...,8)$.

The Gaussian integral (\ref{GaussianInt}) is readily performed. It is given by
the functional determinant of the matrix
\begin{equation}
\mathcal{A}_{xy}^{ij}=\left(
%TCIMACRO{\TeXButton{slash}{\! \not\!} }%
%BeginExpansion
\! \not\!
%EndExpansion
P_{x}^{ij}-M\,\delta^{ij}+g_{w}\,\Gamma^{B}q_{2,x}^{i}\overline{q}_{1,x}%
^{j}\Gamma^{A}\right)  \delta^{(4)}(x-y).\label{Matrix}%
\end{equation}
Colour indices are denoted by $i,j$ $(i,j=1,2,3)$ and spinor indices are
understood. Using the identity $\det\mathcal{A}=\exp tr\ln\mathcal{A}$ and
keeping only the $\mathcal{O}(g_{w})$ term, we obtain the (non-local)
effective Lagrangian $\mathcal{L}_{eff}^{g_{w}}$:
\begin{equation}
\int d^{4}x\mathcal{L}_{eff}^{g_{w}}=-ig_{w}\int d^{4}x\text{ }\overline
{q}_{1,x}^{j}\Gamma^{A}\left(  \frac{1}{%
%TCIMACRO{\TeXButton{slash}{\! \not\!} }%
%BeginExpansion
\! \not\!
%EndExpansion
P-M}\right)  _{xx}^{ji}\Gamma^{B}q_{2,x}^{i}.\label{TraceExplicit}%
\end{equation}
Consequently, the central object needed to compute the local anomalous
operators in request is the propagator$\ $in external field, $\left(
%TCIMACRO{\TeXButton{slash}{\! \not\!}}%
%BeginExpansion
\! \not\!%
%EndExpansion
P-M\right)  _{xy}^{-1}\equiv\left\langle x\left|  \left(
%TCIMACRO{\TeXButton{slash}{\! \not\!}}%
%BeginExpansion
\! \not\!%
%EndExpansion
P-M\right)  ^{-1}\right|  y\right\rangle $, at $x=y$.

We will need the following formal identity:
\begin{equation}
\left(  \frac{1}{%
%TCIMACRO{\TeXButton{slash}{\! \not\!} }%
%BeginExpansion
\! \not\!
%EndExpansion
P-M}\right)  _{xx}=\int\frac{d^{4}p}{\left(  2\pi\right)  ^{4}}\,\frac{1}{%
%TCIMACRO{\TeXButton{slash}{\! \not\!} }%
%BeginExpansion
\! \not\!
%EndExpansion
p+%
%TCIMACRO{\TeXButton{slash}{\! \not\!} }%
%BeginExpansion
\! \not\!
%EndExpansion
P_{x}-M}.\label{ExtFieldProp}%
\end{equation}
Intuitively, it results from the minimal substitution $p_{\mu}\rightarrow
p_{\mu}+P_{\mu}$ in the momentum-space free-quark propagator. The above
expression is in fact properly defined by the infinite series:
\begin{equation}
\left(  \frac{1}{%
%TCIMACRO{\TeXButton{slash}{\! \not\!} }%
%BeginExpansion
\! \not\!
%EndExpansion
P-M}\right)  _{xx}=\sum_{n=0}^{\infty}S_{n}\label{SjSeries}%
\end{equation}
with
\begin{equation}
S_{n}=\int\frac{d^{4}p}{\left(  2\pi\right)  ^{4}}\,\frac{1}{%
%TCIMACRO{\TeXButton{slash}{\! \not\!} }%
%BeginExpansion
\! \not\!
%EndExpansion
p-M}\left(  \left(  -%
%TCIMACRO{\TeXButton{slash}{\! \not\!} }%
%BeginExpansion
\! \not\!
%EndExpansion
P\right)  \frac{1}{%
%TCIMACRO{\TeXButton{slash}{\! \not\!} }%
%BeginExpansion
\! \not\!
%EndExpansion
p-M}\right)  ^{n}.\label{Sj}%
\end{equation}
Let us emphasize that the derivatives in $S_{n}$ will only affect the involved
gluon fields after momentum integration, not the external states $q_{1,2}$.
Indeed, the `plunged' propagator is a function, not an operator.
Eq.(\ref{SjSeries}) will thus eventually turn into a well defined expansion in
$1/M$ provided the gluonic background fluctuations are small.

We will now compute the first few $S_{n}$'s in the gauge-invariant dimensional
regularization scheme. The above integrals can be cast into the form
\begin{equation}
S_{n}=\int\frac{d^{d}p\ \mu^{\varepsilon}}{\left(  2\pi\right)  ^{d}%
}\,\frac{N_{n}}{\left(  p^{2}-M^{2}\right)  ^{n+1}}\label{SjIntegral}%
\end{equation}
with $\varepsilon=4-d$ and $\mu,$ the regularization scale. The numerators
$N_{n}$ are written in a form suitable for integration using exclusively the
Dirac matrices anticommutation relations $\left\{  \gamma_{\mu},\gamma_{\nu
}\right\}  =2g_{\mu\nu}$. In the cases $n=5$ and $n=4$, we also make use of
the following decompositions of $\gamma$-strings on the Clifford basis
\cite{Veltman88}, after momentum integration:
\begin{align}
\gamma_{\mu_{1}}\gamma_{\mu_{2}}\gamma_{\mu_{3}}  & =g_{\mu_{1}\mu_{2}}%
\gamma_{\mu_{3}}-g_{\mu_{1}\mu_{3}}\gamma_{\mu_{2}}+g_{\mu_{2}\mu_{3}}%
\gamma_{\mu_{1}}+i\varepsilon_{\mu_{1}\mu_{2}\mu_{3}\beta}\gamma^{\beta}%
\gamma_{5}\nonumber\\
\gamma_{\mu_{1}}\gamma_{\mu_{2}}\gamma_{\mu_{3}}\gamma_{\mu_{4}}  &
=g_{\mu_{1}\mu_{2}}g_{\mu_{3}\mu_{4}}-g_{\mu_{1}\mu_{3}}g_{\mu_{2}\mu_{4}%
}+g_{\mu_{1}\mu_{4}}g_{\mu_{2}\mu_{3}}-i\varepsilon_{\mu_{1}\mu_{2}\mu_{3}%
\mu_{4}}\gamma_{5}\nonumber\\
& +i\left(  g_{\mu_{1}\mu_{2}}\sigma_{\mu_{4}\mu_{3}}+g_{\mu_{1}\mu_{3}}%
\sigma_{\mu_{2}\mu_{4}}+g_{\mu_{1}\mu_{4}}\sigma_{\mu_{3}\mu_{2}}\right.
\nonumber\\
& \left.  \quad+g_{\mu_{2}\mu_{3}}\sigma_{\mu_{4}\mu_{1}}+g_{\mu_{2}\mu_{4}%
}\sigma_{\mu_{1}\mu_{3}}+g_{\mu_{3}\mu_{4}}\sigma_{\mu_{2}\mu_{1}}\right)
\label{CliffordDecomp}%
\end{align}
where $\gamma_{5}=i\gamma^{0}\gamma^{1}\gamma^{2}\gamma^{3}$, $\sigma_{\mu\nu
}=\frac{i}{2}\left[  \gamma_{\mu},\gamma_{\nu}\right]  $ and $\varepsilon
_{\mu\nu\rho\sigma}$ is the Levi-Civita tensor with $\varepsilon^{0123}=+1$.
These conventions agree with ref.\cite{Peskin}.

Gauge invariance requires $S_{1}$ to vanish. For the next few terms, we easily
obtain:
\begin{align}
S_{2}  & =\frac{-1}{\left(  4\pi\right)  ^{2}}M\left(  N_{\varepsilon}%
-\ln\frac{M^{2}}{\mu^{2}}\right)  P_{\mu}P_{\nu}\sigma^{\mu\nu}\nonumber\\
S_{3}  & =\frac{i}{3\left(  4\pi\right)  ^{2}}\left(  N_{\varepsilon}%
-\ln\frac{M^{2}}{\mu^{2}}\right)  \left[  P^{\nu},\left[  P_{\mu},P_{\nu
}\right]  \right]  \gamma^{\mu}\nonumber\\
S_{4}  & =\frac{-i}{2\left(  4\pi\right)  ^{2}}\frac{1}{M}P_{\mu_{1}}%
P_{\mu_{2}}P_{\mu_{3}}P_{\mu_{4}}\Gamma_{4}^{\mu_{1}\mu_{2}\mu_{3}\mu_{4}%
}\nonumber\\
S_{5}  & =\frac{i}{2\left(  4\pi\right)  ^{2}}\frac{1}{M^{2}}P_{\mu_{1}}%
P_{\mu_{2}}P_{\mu_{3}}P_{\mu_{4}}P_{\mu_{5}}\left(  \Gamma_{5}^{\mu_{1}\mu
_{2}\mu_{3}\mu_{4}\mu_{5}}+\Gamma_{5^{\prime}}^{\mu_{1}\mu_{2}\mu_{3}\mu
_{4}\mu_{5}}\right) \label{SjInt}%
\end{align}
with $N_{\varepsilon}=\frac{2}{\varepsilon}+\ln4\pi-\gamma$, and
\begin{align*}
\Gamma_{4}^{\mu_{1}\mu_{2}\mu_{3}\mu_{4}}  & =-\frac{2}{3}g^{\mu_{1}\mu_{3}%
}g^{\mu_{2}\mu_{4}}+\frac{2}{3}g^{\mu_{1}\mu_{4}}g^{\mu_{2}\mu_{3}%
}-i\varepsilon^{\mu_{1}\mu_{2}\mu_{3}\mu_{4}}\gamma_{5}\\
& +ig^{\mu_{1}\mu_{3}}\sigma^{\mu_{2}\mu_{4}}-ig^{\mu_{2}\mu_{3}}\sigma
^{\mu_{1}\mu_{4}}+ig^{\mu_{2}\mu_{4}}\sigma^{\mu_{1}\mu_{3}}\\
& -\frac{i}{3}g^{\mu_{1}\mu_{2}}\sigma^{\mu_{3}\mu_{4}}-\frac{i}{3}g^{\mu
_{1}\mu_{4}}\sigma^{\mu_{2}\mu_{3}}-\frac{i}{3}g^{\mu_{3}\mu_{4}}\sigma
^{\mu_{1}\mu_{2}}%
\end{align*}
\begin{align*}
\Gamma_{5}^{\mu_{1}\mu_{2}\mu_{3}\mu_{4}\mu_{5}}  & =-\frac{13}{30}g^{\mu
_{2}\mu_{3}}g^{\mu_{4}\mu_{5}}\gamma^{\mu_{1}}+\frac{17}{30}g^{\mu_{2}\mu_{4}%
}g^{\mu_{3}\mu_{5}}\gamma^{\mu_{1}}-\frac{8}{30}g^{\mu_{2}\mu_{5}}g^{\mu
_{3}\mu_{4}}\gamma^{\mu_{1}}\\
& +\frac{17}{30}g^{\mu_{1}\mu_{3}}g^{\mu_{4}\mu_{5}}\gamma^{\mu_{2}}%
-\frac{18}{30}g^{\mu_{1}\mu_{4}}g^{\mu_{3}\mu_{5}}\gamma^{\mu_{2}}%
+\frac{17}{30}g^{\mu_{1}\mu_{5}}g^{\mu_{3}\mu_{4}}\gamma^{\mu_{2}}\\
& -\frac{8}{30}g^{\mu_{1}\mu_{2}}g^{\mu_{4}\mu_{5}}\gamma^{\mu_{3}}%
+\frac{2}{30}g^{\mu_{1}\mu_{4}}g^{\mu_{2}\mu_{5}}\gamma^{\mu_{3}}%
-\frac{18}{30}g^{\mu_{1}\mu_{5}}g^{\mu_{2}\mu_{4}}\gamma^{\mu_{3}}\\
& +\frac{17}{30}g^{\mu_{1}\mu_{2}}g^{\mu_{3}\mu_{5}}\gamma^{\mu_{4}}%
-\frac{18}{30}g^{\mu_{1}\mu_{3}}g^{\mu_{2}\mu_{5}}\gamma^{\mu_{4}}%
+\frac{17}{30}g^{\mu_{1}\mu_{5}}g^{\mu_{2}\mu_{3}}\gamma^{\mu_{4}}\\
& -\frac{13}{30}g^{\mu_{1}\mu_{2}}g^{\mu_{3}\mu_{4}}\gamma^{\mu_{5}}%
+\frac{17}{30}g^{\mu_{1}\mu_{3}}g^{\mu_{2}\mu_{4}}\gamma^{\mu_{5}}%
-\frac{8}{30}g^{\mu_{1}\mu_{4}}g^{\mu_{2}\mu_{3}}\gamma^{\mu_{5}}%
\end{align*}
\begin{align}
\Gamma_{5^{\prime}}^{\mu_{1}\mu_{2}\mu_{3}\mu_{4}\mu_{5}}  & =+\frac{i}%
{3}g^{\mu_{1}\mu_{2}}\varepsilon^{\mu_{3}\mu_{4}\mu_{5}\beta}\gamma_{\beta
}\gamma_{5}-\frac{i}{2}g^{\mu_{1}\mu_{3}}\varepsilon^{\mu_{2}\mu_{4}\mu
_{5}\beta}\gamma_{\beta}\gamma_{5}+\frac{i}{6}g^{\mu_{1}\mu_{4}}%
\varepsilon^{\mu_{2}\mu_{3}\mu_{5}\beta}\gamma_{\beta}\gamma_{5}\nonumber\\
& -\frac{i}{3}g^{\mu_{1}\mu_{5}}\varepsilon^{\mu_{2}\mu_{3}\mu_{4}\beta}%
\gamma_{\beta}\gamma_{5}+\frac{i}{6}g^{\mu_{2}\mu_{3}}\varepsilon^{\mu_{1}%
\mu_{4}\mu_{5}\beta}\gamma_{\beta}\gamma_{5}+\frac{i}{6}g^{\mu_{2}\mu_{5}%
}\varepsilon^{\mu_{1}\mu_{3}\mu_{4}\beta}\gamma_{\beta}\gamma_{5}\nonumber\\
& +\frac{i}{6}g^{\mu_{3}\mu_{4}}\varepsilon^{\mu_{1}\mu_{2}\mu_{5}\beta}%
\gamma_{\beta}\gamma_{5}-\frac{i}{2}g^{\mu_{3}\mu_{5}}\varepsilon^{\mu_{1}%
\mu_{2}\mu_{4}\beta}\gamma_{\beta}\gamma_{5}+\frac{i}{3}g^{\mu_{4}\mu_{5}%
}\varepsilon^{\mu_{1}\mu_{2}\mu_{3}\beta}\gamma_{\beta}\gamma_{5}%
.\label{DiracStructure}%
\end{align}

We now introduce the conventional definitions for the gluon field-strength
tensor $G_{\mu\nu}=\partial_{\mu}A_{\nu}-\partial_{\nu}A_{\mu}-ig_{s}[A_{\mu
},A_{\nu}]$ and its dual $\widetilde{G}_{\mu\nu}=\tfrac{1}{2}\varepsilon
_{\mu\nu\alpha\beta}G^{\alpha\beta}$. After some rearrangements based on the
identities $\left[  P_{\mu},P_{\nu}\right]  =ig_{s}G_{\mu\nu}$ and
$\varepsilon_{\mu\nu\alpha\beta}P^{\alpha}P^{\beta}=ig_{s}\widetilde{G}%
_{\mu\nu}$, we come to the important result:
\begin{align}
S_{2}  & =\frac{-ig_{s}}{2\left(  4\pi\right)  ^{2}}M\left(  N_{\varepsilon
}-\ln\frac{M^{2}}{\mu^{2}}\right)  G_{\mu\nu}\sigma^{\mu\nu}\nonumber\\
S_{3}  & =\frac{-ig_{s}}{3\left(  4\pi\right)  ^{2}}\left(  N_{\varepsilon
}-\ln\frac{M^{2}}{\mu^{2}}\right)  D^{\nu}G_{\mu\nu}\gamma^{\mu}\nonumber\\
S_{4}  & =\frac{-ig_{s}^{2}}{6\left(  4\pi\right)  ^{2}}\frac{1}{M}\left(
G_{\mu\nu}G^{\mu\nu}+\frac{3}{2}iG_{\mu\nu}\widetilde{G}^{\mu\nu}\gamma
_{5}-3iG_{\alpha}^{\mu}G^{\nu\alpha}\sigma_{\mu\nu}-\frac{1}{2g_{s}}D_{\alpha
}D^{\alpha}G_{\mu\nu}\sigma^{\mu\nu}\right) \nonumber\\
S_{5}  & =\frac{-ig_{s}^{2}}{8\left(  4\pi\right)  ^{2}}\frac{1}{M^{2}}\left(
\frac{6}{5}i\left[  D^{\alpha}G_{\mu\alpha},G^{\mu\nu}\right]  \gamma_{\nu
}+\frac{2}{15}i\left[  G_{\mu\alpha},D^{\alpha}G^{\mu\nu}\right]  \gamma_{\nu
}\right. \nonumber\\
& \left.  +\frac{4}{3}\left\{  D^{\alpha}G_{\mu\alpha},\widetilde{G}^{\mu\nu
}\right\}  \gamma_{\nu}\gamma_{5}+\frac{2}{3}\left\{  G_{\mu\alpha},D^{\alpha
}\widetilde{G}^{\mu\nu}\right\}  \gamma_{\nu}\gamma_{5}-\frac{8}{15g_{s}%
}D^{\alpha}D_{\alpha}D^{\nu}G_{\mu\nu}\gamma^{\mu}\right) \label{SjResult}%
\end{align}
with $iD_{\mu}F\equiv\left[  P_{\mu},F\right]  $ for any field $F$
transforming like the $SU(3)_{C}$ adjoint representation.

The dominant penguin-induced corrections to the initial four-quark operator
(\ref{4QuarkOp}) are now explicitly obtained:
\begin{equation}
\mathcal{L}_{eff}^{g_{w}}=-ig_{w}\sum_{n=2}^{5}\overline{q}_{1}\Gamma^{A}%
S_{n}\Gamma^{B}q_{2}+\mathcal{O}\left(  1/M^{3}\right)  .\label{j+3dLagr}%
\end{equation}
Remarkably, trace and axial anomalous operators are in principle already
generated at the lowest possible order in the momentum expansion, as inferred
from the colour-singlet part of the scalar and pseudoscalar gluon densities
contained in $S_{4}^{ji}.$

Let us now apply these results to the case of the standard-model
short-distance weak operator evolution.

\section{Anomalies from standard-model operator evolution}

Tree-level integration of the $W$ gauge boson in $V-A$ charged current
interactions leads to effective operators of the type (\ref{4QuarkOp}) with
$\Gamma^{A(B)}=\gamma_{\mu}^{(\mu)}\left(  1-\gamma_{5}\right)  $.

Let us first focus on the well-suited case of charm quark loop contributions
to Kaon decays. The effective coupling associated with the tree-level operator
defined at the $M_{W}$ scale
\begin{equation}
Q_{2}^{(c)}=4(\overline{d}_{L}\gamma_{\mu}c_{L})(\overline{c}_{L}\gamma^{\mu
}s_{L})\label{FermiOp}%
\end{equation}
is given by
\begin{equation}
g_{w}^{SM}=-V_{cs}V_{cd}^{*}G_{F}/\sqrt{2},\label{WeakCoupling}%
\end{equation}
with the notation $q_{L}\equiv\tfrac{1}{2}\left(  1-\gamma_{5}\right)  q.$

The chiral structure of this operator obviously prevents the even $n$
contributions from appearing. The first non-vanishing terms are thus obtained
for $n=3$ and $n=5$. They correspond to dimension-six and -eight operators,
respectively:
\begin{equation}
\mathcal{L}_{eff}^{\Delta S=1}=\mathcal{L}_{6d}+\mathcal{L}_{8d}%
+...\label{LeffExpSM}%
\end{equation}
Still, it is worthwhile to analyse them since, as we shall see, they can also
carry some strong anomaly effects.

\subsection*{Dimension-six operators}

The dominant operators in the $m_{c}^{-1}$ expansion (\ref{LeffExpSM}) have
been computed a long time ago \cite{6dPenguins}. Yet, we find it useful to
illustrate how the external field formalism works in a simple case. It also
allows us to fix our conventions. The effect of incomplete GIM mechanism
\cite{Bardeen87} above $m_{c}$ is at the next-to-leading level \cite{Buras93},
i.e. beyond the scope of this paper. So, the leading short-distance evolution
of $Q_{2}^{(c)}$ from $M_{W}$ down to $m_{c}$ only involves the usual
current-current operators \cite{Gaillard74}, which disappear once the charm
quark is integrated out. Inserting the explicit form of $S_{3}$ in
eq.(\ref{j+3dLagr}), we obtain then at the leading order in the strong
coupling constant:
\begin{equation}
\mathcal{L}_{6d}=\frac{G_{F}}{\sqrt{2}}V_{cs}V_{cd}^{*}\frac{g_{s}}{6\pi^{2}%
}\ln\left(  \frac{m_{c}^{2}}{\mu^{2}}\right)  \overline{d}_{L}D^{\nu}G_{\mu
\nu}\gamma^{\mu}s_{L}\label{6dLagr}%
\end{equation}
with $\mu<m_{c}$. Recall that the standard computation of dimension-six
gluonic penguin operators involves the use of the classical equations of
motion:
\begin{equation}
D^{\nu}G_{\mu\nu}=g_{s}\sum_{q=u,d,s}\left(  \overline{q}\gamma_{\mu
}\frac{\lambda^{a}}{2}q\right)  \frac{\lambda^{a}}{2}.\label{EqMotionGluon}%
\end{equation}
Applying these equations on Lagrangian (\ref{6dLagr}) together with the
identity $\left(  \lambda^{a}\right)  _{ij}\left(  \lambda^{a}\right)
_{kl}=2\left(  \delta_{il}\delta_{kj}-\delta_{ij}\delta_{kl}/3\right)  $ and
the relevant Fierz reorderings, we indeed recover the well-known
current-current and density-density colour-singlet operators:
\begin{equation}
\mathcal{L}_{6d}=\frac{G_{F}}{\sqrt{2}}V_{cs}V_{cd}^{*}\frac{\alpha_{s}}{4\pi
}\ln\left(  \frac{m_{c}^{2}}{\mu^{2}}\right)  \sum_{k=3}^{6}c_{k}^{(6)}%
Q_{k}^{(6)}\label{PenguinLagr}%
\end{equation}
with
\[
\begin{tabular}
[c]{ll}%
$Q_{3}^{(6)}=4(\overline{d}_{L}\gamma_{\mu}s_{L})(\overline{q}_{L}\gamma^{\mu
}q_{L})\quad$ & $Q_{5}^{(6)}=4(\overline{d}_{L}\gamma_{\mu}s_{L})(\overline
{q}_{R}\gamma^{\mu}q_{R})$\\
$Q_{4}^{(6)}=4(\overline{d}_{L}\gamma_{\mu}q_{L})(\overline{q}_{L}\gamma^{\mu
}s_{L})\quad$ & $Q_{6}^{(6)}=-8(\overline{d}_{L}q_{R})(\overline{q}_{R}s_{L})$%
\end{tabular}
\]
and
\[
c_{3}^{(6)}=c_{5}^{(6)}=-\frac{1}{9},\quad c_{4}^{(6)}=c_{6}^{(6)}=\frac{1}%
{3}.
\]
In our conventions, $q_{R}\equiv\tfrac{1}{2}\left(  1+\gamma_{5}\right)  q$
and the light flavours $q=u,d,s$ are summed over.

\subsection*{Dimension-eight operators}

The dimension-eight penguin-like operators are most easily obtained from the
contribution of $S_{5}$ to eq.(\ref{j+3dLagr}), which gives:
\begin{equation}
\mathcal{L}_{8d}=-\frac{G_{F}}{\sqrt{2}}V_{cs}V_{cd}^{*}\frac{\alpha_{s}}%
{4\pi}\frac{1}{m_{c}^{2}}\sum_{k=1}^{5}c_{k}^{(8)}Q_{k}^{(8)}\label{8dLagr}%
\end{equation}
with
\[
\begin{tabular}
[c]{l}%
$\smallskip Q_{1}^{(8)}=i\overline{d}_{L}\left[  D^{\alpha}G_{\mu\alpha
},G^{\mu\nu}\right]  \gamma_{\nu}s_{L}$\\
$\smallskip Q_{2}^{(8)}=i\overline{d}_{L}\left[  G_{\mu\alpha},D^{\alpha
}G^{\mu\nu}\right]  \gamma_{\nu}s_{L}$\\
$\smallskip Q_{3}^{(8)}=\overline{d}_{L}\left\{  D^{\alpha}G_{\mu\alpha
},\widetilde{G}^{\mu\nu}\right\}  \gamma_{\nu}s_{L}$\\
$\smallskip Q_{4}^{(8)}=\overline{d}_{L}\left\{  G_{\mu\alpha},D^{\alpha
}\widetilde{G}^{\mu\nu}\right\}  \gamma_{\nu}s_{L}$\\
$Q_{5}^{(8)}=\overline{d}_{L}D^{\alpha}D_{\alpha}D^{\nu}G_{\mu\nu}\gamma^{\mu
}s_{L}$%
\end{tabular}
\]
and
\[
c_{1}^{(8)}=\frac{6}{5},\ c_{2}^{(8)}=\frac{2}{15},\ c_{3}^{(8)}=-\frac{4}%
{3},\ c_{4}^{(8)}=-\frac{2}{3},\ c_{5}^{(8)}=-\frac{8}{15}g_{s}^{-1}.
\]
It may be remarked that this effective Lagrangian is invariant under the $CPS
$ symmetry, as it should \cite{Bernard85}.

Dimension-eight operators for Kaon decays have already been considered by the
authors of ref.\cite{PenPiv1}. Yet, their early use of the equations of motion
introduces an explicit dependence on the light quark masses $m_{d,s}$, which
is artificial as the `plunged' propagator is fundamentally a function, not a
differential operator acting on external spinors. Besides, their ensuing
truncated expansion in $m_{d,s}$ is not allowed.

Notice that the last operator in eq.(\ref{8dLagr}) is still linear in the
gluon field-strength tensor. As such, it contributes to the finite part of the
standard $\overline{d}sg$ one-loop diagram, providing thus an interesting
check of our calculation. Indeed, the gluon momentum expansion of the `vacuum
polarization' integral that naturally arises from the four-quark
approximation
\begin{equation}
\int_{0}^{1}dx\text{\thinspace}x(1-x)\ln\left(  \frac{m_{c}^{2}-x(1-x)q^{2}%
}{\mu^{2}}\right)  =\frac{1}{6}\left(  \ln\frac{m_{c}^{2}}{\mu^{2}}%
-\frac{1}{5}\frac{q^{2}}{m_{c}^{2}}\right)  +\mathcal{O}(q^{4}/m_{c}%
^{4})\label{PenguinFF}%
\end{equation}
displays the correct relative weight between the operators of eq.(\ref{6dLagr}%
) and eq.(\ref{8dLagr}) with $k=5$.

Let us now investigate the anomalous content of Lagrangian (\ref{8dLagr}).
Operators possibly involving $G_{a}^{\mu\nu}G_{\mu\nu}^{a}$ and $G_{a}^{\mu
\nu}\widetilde{G}_{\mu\nu}^{a}$ gluon densities are necessarily contained in
the combinations
\begin{equation}
Q_{1}^{(8)}+Q_{2}^{(8)}=i\overline{d}_{L}D^{\alpha}\left[  G_{\mu\alpha
},G^{\mu\nu}\right]  \gamma_{\nu}s_{L}\label{CombiS5TA}%
\end{equation}
and
\begin{equation}
Q_{3}^{(8)}+Q_{4}^{(8)}=\overline{d}_{L}D^{\alpha}\left\{  G_{\mu\alpha
},\widetilde{G}^{\mu\nu}\right\}  \gamma_{\nu}s_{L},\label{CombiS5}%
\end{equation}
respectively.

Remarkably, the gluonic part of the first operator is a pure $SU(3)_{C}$
octet, and consequently cannot carry any trace anomalous effect, in
contradistinction to the results of ref.\cite{PenPiv1}.

On the other hand, the second operator does contain a $SU(3)_{C}$ singlet
part. Indeed, dropping a harmless total derivative, eq.(\ref{CombiS5}) can be
rewritten:
\begin{equation}
Q_{3}^{(8)}+Q_{4}^{(8)}=\left[  -i\overline{d}_{L}\overleftarrow{P}%
^{\dagger\alpha}\gamma_{\nu}\left\{  \frac{\lambda^{a}}{2},\frac{\lambda^{b}%
}{2}\right\}  s_{L}+i\overline{d}_{L}\left\{  \frac{\lambda^{a}}%
{2},\frac{\lambda^{b}}{2}\right\}  \gamma_{\nu}P^{\alpha}s_{L}\right]
G_{\mu\alpha}^{a}\widetilde{G}^{\mu\nu,b}\label{CombiS5AA}%
\end{equation}
with $\overleftarrow{P}_{\alpha}^{\dagger}=-i\overleftarrow{\partial}_{\alpha
}+g_{s}A_{\alpha}$, the derivative acting on the left. Taking then the trace
over colour and Lorentz indices, which amounts to the substitution
\begin{equation}
G_{\mu\alpha}^{a}\widetilde{G}^{\mu\nu,b}\rightarrow\frac{\delta^{ab}}%
{8}\frac{\delta_{\alpha}^{\nu}}{4}G_{\rho\sigma}^{c}\widetilde{G}_{c}%
^{\rho\sigma},\label{ColLoTr}%
\end{equation}
and including the coefficients $c_{3}^{(8)}\ $and $c_{4}^{(8)}$, we come to
the following axial anomalous operator:
\begin{equation}
Q_{AA}^{(8)}=\frac{-i}{12}\left[  \overline{d}_{L}%
%TCIMACRO{\TeXButton{slash}{\! \not\!}}%
%BeginExpansion
\! \not\!%
%EndExpansion
Ps_{L}-\overline{d}_{L}%
%TCIMACRO{\TeXButton{slash}{\! \not\!\!}}%
%BeginExpansion
\! \not\!\!%
%EndExpansion
\overleftarrow{P}^{\dagger}s_{L}\right]  G_{\mu\nu}^{a}\widetilde{G}_{a}%
^{\mu\nu}.\label{AAOp}%
\end{equation}
The quark classical equations of motion
\begin{equation}%
%TCIMACRO{\TeXButton{slash}{\! \not\!}}%
%BeginExpansion
\! \not\!%
%EndExpansion
Ps=m_{s}s,\quad\overline{d}%
%TCIMACRO{\TeXButton{slash}{\! \not\!\!}}%
%BeginExpansion
\! \not\!\!%
%EndExpansion
\overleftarrow{P}^{\dagger}=m_{d}\overline{d}\label{EqMotionQuark}%
\end{equation}
may eventually be used to put this result in a more conventional form. Yet,
the matrix element of $G_{\mu\nu}^{a}\widetilde{G}_{a}^{\mu\nu}$ between the
vacuum and a neutral pion state requires isospin violation in the
factorization approximation. The charm-induced axial anomaly effect on
hadronic $K$ decays is thus negligible as far as the $\Delta I=1/2$ amplitude
is concerned.

\subsection*{Dimension-five and -seven operators}

We have seen that the chiral structure of the tree-level operator
(\ref{FermiOp}) prevents any $S_{2n}$ contribution from surviving at the
effective Lagrangian level. However, it is well-known that dimension-five
`chromo-magnetic' penguin operators do arise in the Standard Model once we go
beyond the four-quark approximation. A full short-distance calculation with
massive $W$ propagation gives indeed for both the charm and top contributions
\cite{dsg}:
\begin{equation}
\mathcal{L}_{5d}=\frac{G_{F}}{\sqrt{2}}V_{q^{\prime}s}V_{q^{\prime}d}%
^{*}\frac{g_{s}}{32\pi^{2}}c_{q^{\prime}}^{(5)}Q^{(5)}\text{\quad}(q^{\prime
}=c,t)\label{ChromoPeng}%
\end{equation}
with
\[
Q^{(5)}=m_{s}\overline{d}_{L}G_{\mu\nu}\sigma^{\mu\nu}s_{R}+m_{d}\overline
{d}_{R}G_{\mu\nu}\sigma^{\mu\nu}s_{L}
\]
and
\[
c_{q^{\prime}}^{(5)}(m_{q^{\prime}}\ll M_{W})=\mathcal{O}\left(
\frac{m_{q^{\prime}}^{2}}{M_{W}^{2}}\right)  \text{,\quad}c_{q^{\prime}}%
^{(5)}(m_{q^{\prime}}\gg M_{W})=\mathcal{O}\left(  1\right)  .
\]
So, the longitudinal part of the $W$ propagator induces a chiral structure
compatible with the $S_{2n}$ operators once the light quark equations of
motion are used. However, the price to pay is a suppression factor of the
order of $m_{s}m_{c}/M_{W}^{2}$ or $m_{s}/m_{t}$. This can be seen explicitly
from eqs.(\ref{ChromoPeng}) and (\ref{SjResult}) in the case of $S_{2}$.
Still, similar suppressions are expected to occur also for higher $S_{2n}$
structures in hadronic $K$ decays. Indeed, the helicity flip undergone by
external spinors implies a linear dependence on the light quark masses, while
the denominators can be inferred from the limit $M_{W}\rightarrow\infty$ in
the charm case and from dimensional arguments in the top case.

Let us now turn to hadronic $B$ decays. In exactly the same way, the virtual
top quark contribution is suppressed by a factor of the order of $m_{b}/m_{t}
$. On the other hand, the charm loop contribution induced by the tree-level
operator $Q_{2}^{(c)}=4(\overline{s}_{L}\gamma_{\mu}c_{L})(\overline{c}%
_{L}\gamma^{\mu}b_{L})$ is rather troublesome since the momenta associated
with gluon fields can now be larger than $m_{c}$. For illustration, the
time-like squared momentum of the single gluon configuration generated by the
`vacuum polarization' integral (\ref{PenguinFF}) is likely in the range of
$m_{b}^{2}/4\lesssim q^{2}\lesssim m_{b}^{2}/2$ for two-body exclusive $B$
decays \cite{Gérard91}. If such is the case, the expansion in inverse powers
of the charm mass obviously breaks down, non-local operators survive and an
absorptive component may even arise for $q^{2}\geq4m_{c}^{2}$. Consequently,
we cannot exclude non-negligible anomalous effects associated with $S_{2n+1}$
chiral structures in $B\rightarrow\eta^{\prime}K$, unless the factorization
approximation is called for help.

\subsection*{Factorized amplitudes}

The factorized part of the $B\rightarrow\eta^{\prime}K$ decay amplitude
reads:
\begin{equation}
\left\langle \eta^{\prime}K\left|  Q_{2}^{(c)}\right|  B\right\rangle
_{F}=-\frac{1}{3}\left\langle \eta^{\prime}\left|  \overline{c}\gamma_{\mu
}\gamma_{5}c\right|  0\right\rangle \left\langle K\left|  \overline{s}%
\gamma^{\mu}b\right|  B\right\rangle .\label{FactoBKEtap}%
\end{equation}
The fact that heavy quarks can only contribute to light pseudoscalar decay
constants when propagating in a loop implies:
\begin{equation}
\left\langle \eta^{\prime}\left|  \overline{c}\gamma^{\mu}\gamma_{5}c\right|
0\right\rangle =\left\langle \eta^{\prime}\left|  -iTr\left[  \left(
%TCIMACRO{\TeXButton{slash}{\! \not\!}}%
%BeginExpansion
\! \not\!%
%EndExpansion
P-m_{c}\right)  _{xx}^{-1}\gamma^{\mu}\gamma_{5}\right]  \right|
0\right\rangle ,\label{EtaMatrElGen}%
\end{equation}
with the trace taken over both spinor and colour indices. Inserting the first
few terms of the charm quark propagator expansion (\ref{SjResult}) in the
above matrix element, we come to the conclusion that only part of $S_{5}$
survives to give:
\begin{equation}
\left\langle \eta^{\prime}\left|  \overline{c}\gamma^{\mu}\gamma_{5}c\right|
0\right\rangle =\left\langle \eta^{\prime}\left|  \frac{\alpha_{s}}{12\pi
}\frac{1}{m_{c}^{2}}\left[  2\left(  D^{\alpha}G_{\beta\alpha}\right)
^{a}\widetilde{G}_{a}^{\beta\mu}+G_{\beta\alpha}^{a}\left(  D^{\alpha
}\widetilde{G}^{\beta\mu}\right)  _{a}\right]  \right|  0\right\rangle
+\mathcal{O}\left(  1/m_{c}^{4}\right)  .\label{EtaMatrEl}%
\end{equation}
While this quantity remains to be estimated, it is already interesting to
notice that the hadronization of the $\overline{c}c$ pair into the
$\eta^{\prime}$ meson proceeds indeed through $S_{2n+1}$ gluon structures,
unlike what emerges from ref.\cite{Halperin97} where the charm quark equation
of motion is used from the start.

The axial anomaly part of eq.(\ref{EtaMatrEl}) is given by:
\begin{equation}
\left\langle \eta^{\prime}\left|  \overline{c}\gamma^{\mu}\gamma_{5}c\right|
0\right\rangle _{AA}=\frac{1}{32m_{c}^{2}}\left\langle \eta^{\prime}\left|
\frac{\alpha_{s}}{\pi}\partial^{\mu}\left(  G_{\alpha\beta}^{a}\widetilde
{G}_{a}^{\alpha\beta}\right)  \right|  0\right\rangle \equiv-if_{\eta^{\prime
}}^{(c)AA}p_{\eta^{\prime}}^{\mu}\label{EtaMatrElAA}%
\end{equation}
with
\begin{equation}
f_{\eta^{\prime}}^{(c)AA}=\frac{-1}{32m_{c}^{2}}\left\langle \eta^{\prime
}\left|  \frac{\alpha_{s}}{\pi}G_{\alpha\beta}^{a}\widetilde{G}_{a}%
^{\alpha\beta}\right|  0\right\rangle \simeq-2\text{ }MeV.\label{AAFormFactor}%
\end{equation}
The $QCD$ sum rules calculation of ref.\cite{Novikov79} has been used to
estimate the last matrix element. Consequently, the charm quark induced axial
anomaly can only account for a few per cent of the experimental $B\rightarrow
\eta^{\prime}K$ decay amplitude in the factorization approximation.

Note that charm quark loop effects in factorized $B\rightarrow f_{0}(\sigma)K$
decay amplitudes may be treated in a similar way. Yet these appear to be even
more suppressed:
\begin{equation}
\left\langle f_{0}(\sigma)\left|  \overline{c}\gamma^{\mu}c\right|
0\right\rangle =\left\langle f_{0}(\sigma)\left|  -iTr\left[  \left(
%TCIMACRO{\TeXButton{slash}{\! \not\!}}%
%BeginExpansion
\! \not\!%
%EndExpansion
P-m_{c}\right)  _{xx}^{-1}\gamma^{\mu}\right]  \right|  0\right\rangle
=\mathcal{O}\left(  1/m_{c}^{4}\right)  .\label{PiPiMatrEl}%
\end{equation}

\section{Anomalies beyond standard-model operator evolution}

The minimal non-standard operator leading to $S_{2n}$-related anomalous
effects reads
\begin{equation}
Q=(\overline{d}_{L}c_{R})(\overline{c}_{L}s_{R})\label{DensityOp}%
\end{equation}
in the case of hadronic $K$ decays. This tree-level operator generically
arises in multi-Higgs models \cite{Branco85}. In this framework, the following
dimension-seven operators are directly obtained from $S_{4}$:
\begin{equation}
\mathcal{L}_{7d}=-g_{w}\frac{\alpha_{s}}{24\pi}\frac{1}{m_{c}}\sum_{k=1}%
^{4}c_{k}^{(7)}Q_{k}^{(7)}\label{7dLagr}%
\end{equation}
with
\[
\begin{tabular}
[c]{l}%
$\smallskip Q_{1}^{(7)}=\overline{d}_{L}G_{\mu\nu}G^{\mu\nu}s_{R}$\\
$\smallskip Q_{2}^{(7)}=i\overline{d}_{L}G_{\mu\nu}\widetilde{G}^{\mu\nu}%
s_{R}$\\
$\smallskip Q_{3}^{(7)}=i\overline{d}_{L}G_{\alpha}^{\mu}G^{\nu\alpha}%
\sigma_{\mu\nu}s_{R}$\\
$\smallskip Q_{4}^{(7)}=\overline{d}_{L}D_{\alpha}D^{\alpha}G_{\mu\nu}%
\sigma^{\mu\nu}s_{R}$%
\end{tabular}
\]
and
\[
c_{1}^{(7)}=1,\text{ }c_{2}^{(7)}=\frac{3}{2},\text{ }c_{3}^{(7)}=-3,\text{
}c_{4}^{(7)}=\frac{-1}{2}g_{s}^{-1}.
\]
In particular, their anomalous content is given by:
\begin{equation}
Q_{AA+TA}^{(7)}=\frac{1}{6}\overline{d}_{L}s_{R}\left(  G_{\mu\nu}^{a}%
G_{a}^{\mu\nu}+\frac{3}{2}iG_{\mu\nu}^{a}\widetilde{G}_{a}^{\mu\nu}\right)
\label{AAplusTAOp}%
\end{equation}
if the coefficients $c_{1}^{(7)}$ and $c_{2}^{(7)}$ are included. However, the
effective coupling constant $g_{w}$ is in principle suppressed by the
characteristic scale of the new physics involved.

In the factorization approximation, the trace and axial anomalous operators
merged in eq.(\ref{AAplusTAOp}) are the only contributions of eq.(\ref{7dLagr}%
) to hadronic $K$ decay amplitudes. They can thus be directly inferred from
the factorization of $Q$ matrix elements. Indeed, the charm quark loop
contribution in
\begin{align}
\left\langle \pi^{+}\pi^{-}\left|  Q\right|  \overline{K}^{0}\right\rangle
_{F}  & =\frac{1}{12}\left\langle \pi^{+}\pi^{-}\left|  \overline{c}c\right|
0\right\rangle \left\langle 0\left|  \overline{d}_{L}s_{R}\right|
\overline{K}^{0}\right\rangle \label{FactoKPiPi}\\
\left\langle \pi^{-}\pi^{0}\left|  Q\right|  K^{-}\right\rangle _{F}  &
=\frac{1}{12}\left\langle \pi^{0}\left|  \overline{c}\gamma_{5}c\right|
0\right\rangle \left\langle \pi^{-}\left|  \overline{d}_{L}s_{R}\right|
K^{-}\right\rangle \label{FactoKPiPi0}%
\end{align}
is readily estimated using the `plunged' propagator, provided we make the
replacement $\gamma^{\mu}\rightarrow1\ $in eq.(\ref{PiPiMatrEl}) and
$\gamma^{\mu}\gamma_{5}\rightarrow\gamma_{5}$ in eq.(\ref{EtaMatrElGen}). Let
us point out that this simply amounts to the well-known substitutions
\cite{Shifman78}:
\begin{align}
m_{c}\overline{c}c  & \rightarrow\frac{-2}{3}\frac{\alpha_{s}}{8\pi}G_{\mu\nu
}^{a}G_{a}^{\mu\nu}+\mathcal{O}(\mu^{2}/m_{c}^{2})\label{DecouplTrace}\\
m_{c}\overline{c}i\gamma_{5}c  & \rightarrow\frac{\alpha_{s}}{8\pi}G_{\mu\nu
}^{a}\widetilde{G}_{a}^{\mu\nu}+\mathcal{O}(\mu^{2}/m_{c}^{2}%
)\label{DecouplAxial}%
\end{align}
for heavy quark contributions to light hadron matrix elements, providing thus
another check of our central result (\ref{SjResult}).

\section{Conclusion}

We have proposed a simple way to systematically generate penguin-like
operators from the inverse mass expansion of the heavy quark propagator
plunged in $QCD$ external fields. This expansion has been explicitly performed
up to the fifth order. The result, expressed in a compact and comprehensible
form, is also suitable for other applications like, as we have seen, the heavy
quark contribution to light meson decay constants.

This formalism has allowed us to clarify the role of the strong axial and
trace anomalous operators in hadronic weak decays involving $c,$ $b$ or $t$
quark loops. Our results may be summarized as follows:\newline $-$ The virtual
$c$ contribution to $K$ decay amplitudes provides us with an ideal laboratory
to study the trace anomaly effects since the scalar gluon density could softly
convert into a pair of light pions. However, in the Standard Model, the
short-distance rise of the dimension-seven trace and axial anomalous operators
is necessarily $1/M_{W}^{2}$-suppressed due to the chiral structure of the
four-quark effective weak interactions, while it is shown that there is no
trace anomaly effect from dimension-eight operators. Besides, from
factorization arguments, we conjecture a small impact of $c$ quark induced
strong anomalies on hadronic $B$ decays. Notice that a complementary
conclusion has been reached recently by the authors of ref.\cite{Eeg03} for
the axial `anomaly tail' due to highly virtual gluons, i.e. above the $m_{b}$
scale.\newline $-$ The formalism applies to $b$ quark loop effects in $D$
meson decays too. However, these suffer in any case from a multi-Cabibbo
suppression.\newline $-$ The $t$ quark induced anomalous operators, on the
other hand, require to go beyond the four-quark approximation. Yet, a simple
dimensional argument leads us to conclude that the chiral suppression in
$K(B)$ decays is now controlled by the $m_{s(b)}/m_{t}$ ratio.

Consequently, if the strong axial and trace anomalies have a role to play in
hadronic weak decays, in particular in $B\rightarrow\eta^{\prime}K_{S}$ and
$K\rightarrow\pi\pi$, they should arise from either nonperturbative effects
induced by the light $s,$ $d$ and $u$ quarks or new physics beyond the
Standard Model.

Our formalism also provides us with a systematic classification of
charm-induced operators for $K$ decays. In particular, a complete set of
gauge-invariant dimension-eight operators is displayed. Their effect, of the
order of $m_{K}^{2}/m_{c}^{2}$, might compete with other non-leading
short-distance $QCD$ corrections \cite{Buras93}. As such, they obviously
deserve further attention, independently of their anomalous content. Let us
notice that distinct $1/\mu^{2}$ effects beyond dimensional regularization
have been discussed in ref.\cite{Cirigliano00}.

Finally, our method is readily extended to account for $c$ quark contributions
to radiative $K$ decays. Indeed, the inclusion of photons simply proceeds
trough the substitutions:
\begin{align*}
g_{s}G_{\mu\nu}  & \rightarrow g_{s}G_{\mu\nu}+eqF_{\mu\nu}\\
g_{s}D_{\alpha_{1}}...D_{\alpha_{n}}G_{\mu\nu}  & \rightarrow g_{s}%
D_{\alpha_{1}}...D_{\alpha_{n}}G_{\mu\nu}+eq\partial_{\alpha_{1}}%
...\partial_{\alpha_{n}}F_{\mu\nu}%
\end{align*}
with $F^{\mu\nu}$, the photon field-strength tensor and $q$, the heavy quark
charge in units of the electron charge $e$.

\section*{Acknowledgments}

We are pleased to thank C. Smith and J. Weyers for stimulating discussions and
useful comments. This work was supported by the Belgian Science Policy through
the Interuniversity Attraction Pole P5/27. S. Trine also acknowledges
financial support from IISN.

\end{document}